\documentstyle[12pt,epsfig]{article}

\newcommand{\be}{\begin{eqnarray}}
\newcommand{\ee}{\end{eqnarray}}
\newcommand{\e}{\epsilon}
\renewcommand{\l}{\lambda}
\renewcommand{\Re}{{\rm Re}}
\renewcommand{\Im}{{\rm Im}}

\begin{document}

\newpage
\setcounter{page}{0}

\begin{titlepage}
 \begin{flushright}
 \hfill{\bf (hep-ph/9904470)}\\
 \hfill{YUMS 99--010}\\

 \end{flushright}
\vspace*{0.8cm}

\begin{center}
{\large\bf CP Violation in the Semileptonic $B_{l4}$ 
($B^\pm \rightarrow \pi^+ \pi^- l^\pm \nu$) 
 Decays}
\end{center}
\vskip 0.8cm
\begin{center}
{\sc C. S.~Kim$^{\mathrm{a,}}$\footnote{e-mail : kim@cskim.yonsei.ac.kr,
~~ http://phya.yonsei.ac.kr/\~{}cskim/},
 Jake Lee$^{\mathrm{a,}}$\footnote{e-mail : jilee@theory.yonsei.ac.kr} and 
 W.~Namgung$^{\mathrm{b,}}$\footnote{e-mail : ngw@cakra.dongguk.ac.kr}}

\vskip 0.5cm

\begin{small} 
$^{\mathrm{a}}$ {\it Department of Physics, Yonsei University 120-749, Seoul, 
                  Korea} \\
\vskip 0.2cm
$^{\mathrm{b}}$ {\it Department of Physics, Dongguk University 100-715, 
                  Seoul, Korea}
\end{small}
\end{center}

\vspace{0.5cm}
\begin{center}
 (\today)
\end{center}

\setcounter{footnote}{0}
\begin{abstract}

\noindent
Direct CP violations in $B_{l4}$ decays ($B^\pm \to \pi^+\pi^- l^\pm {\nu}_l$) 
are investigated within the Standard Model (SM) and also in its extensions. 
In the decay processes, we include various excited states as intermediate states 
decaying to the final hadrons, $\pi^+ + \pi^-$. 
The CP violation within the SM is induced by the interferences
between intermediate resonances with different quark flavors.
As extensions of the SM, we consider CP violations implemented through
complex scalar-fermion couplings in the multi-Higgs doublet model 
and the scalar-leptoquark models. 
We calculate the CP-odd rate asymmetry and the optimal asymmetry.
We find that the optimal asymmetry 
can be measured at $1\sigma$ level with about 
$10^9$ $B$-meson pairs in the SM case and $10^3$--$10^7$ pairs in
the extended model case, 
for maximally-allowed values of CP-odd parameters in each case.
\vskip 1cm
PACS numbers: 11.30.Er, 13.20.Hw
\end{abstract}
\vskip 1cm
%
\end{titlepage}

\newpage
\baselineskip .29in
\renewcommand{\thefootnote}{\alph{footnote}}

\section{Introduction}

\noindent
Semileptonic 4-body decays of $B$-mesons with emission of a single pion
have been studied in detail by many authors \cite{BL,korner,bl4}.
Recently we investigated the possibility of probing direct CP violation in
the decay $B^\pm\to D\pi l^\pm \nu$ \cite{bl4} 
in extensions of the Standard Model (SM), where we extended the weak 
charged current by including a scalar-exchange interaction 
with a complex coupling, and considered as specific models the
multi-Higgs doublet (MHD) model and the scalar-leptoquark (SLQ) models.
In the present work, we investigate the same possibility in the decay of 
$B^\pm\to (\pi^+\pi^-) l^\pm \nu$. In this case we find there may be 
direct CP violation even within the SM.

As is well known, in order to observe direct CP violation effects, there should
exist interferences not only through weak CP-violating phases but also with 
different CP-conserving strong phases.
In the decay of $B^\pm\to \pi\pi l^\pm \nu$, we consider it as a two-stage process: 
$B\to (\sum_{i} M_i\to \pi\pi)l\nu$, where $M_i$ stands for an intermediate
state which is decaying to $\pi^+ + \pi^-$.
In this picture the CP-conserving phases may come from the absorptive parts
of the intermediate resonances.
Here we try to include as many as possible intermediate states decaying 
to $\pi^+ + \pi^-$,
so that they could represent a pseudo complete set of the relevant decay.
The candidates in $b\to u$ transition are $\rho$, $f_0$ and $f_2$ mesons,
which decay dominantly to $2\pi$ mode (See Table.~1).
Furthermore, we find that even in $b\to c$ transition 
a $D^0$ meson can decay to $\pi^++\pi^-$,
although its branching fraction is very small compared to those of 
$u\bar{u}$  states.
However, we can  find that the contribution through an intermediate $D$ 
meson is not negligible because of CKM favored nature of $b\to c$ transition
compared to the $b\to u$ one of $u\bar{u}$ states, $\rho$, $f_0$ and $f_2$
mesons.
If we include $D$ meson as an intermediate state as well as the $u\bar{u}$
states, direct CP violation may arise even within the SM through their relative
weak phases of the different CKM matrix elements ($V_{cb}$ and $V_{ub}$).
Therefore, we first consider CP violation within the SM by including
$\rho$, $f_0$, $f_2$ and $D$ mesons\footnote{
Here we are including fully known resonances only, 
and neglecting  possible non-resonant $2\pi$ decays.
A significant experimental enhancement can be made, if we use the reduced 
kinematic region around 
1.4 GeV $\leq \sqrt{(p_\pi+p_\pi)^2} \leq$ 1.9 GeV (See Table.~1).}
as intermediate states decaying to $\pi^+\pi^-$.
Next we also consider CP violations in extensions of the SM,
in which we use a cutoff to the final state $\pi\pi$ invariant mass
so that the effects of $D$ meson cannot enter, thus ensuring
that the result is solely from new physics.

In Section 2, we present our formalism dealing with $B\to \pi\pi l \nu$ decays
within the SM and in its extensions, 
and the observable asymmetries are considered in Section 3.
Section 4 contains our numerical results and conclusions.
Presented in Appendix are all the relevant formulae we use here.

\begin{table}
{Table~1}. {Properties and branching ratios of $\pi^+\pi^-$ resonances}\par
\begin{tabular}{|c|l|c|c|c|c|}
\hline
Label $i$ & $\qquad\quad M_i$ & $J^P$ & $m_i$ (MeV) & $\Gamma_i$ (MeV) &
${\cal BR}_i(M_i\to \pi^+ \pi^-)$\\
\hline
$0$ & $M_0=f_0(980)$ & $0^+$ & $980$ & $40\sim 100$ & 0.52\\
$0^\prime$ & $M_{0^\prime}=f_0(1500)$ & $0^+$ & $1500$ & $112$ & 0.3\\
$1$ & $M_1=\rho (770)$ & $1^-$ & $770$ & $151$ & 1\\
$1^\prime$ & $M_{1^\prime}=\rho (1700)$ & $1^-$ & $1700$ & $240$ & 0.3\\
$2$ & $M_2=f_2 (1270)$ & $2^+$ & $1275$ & $186$ & 0.56\\
$3$ & $M_3=D^0$ & $0^-$ & $1865$ & $1.59\times 10^{-9}$ & $1.53\times 10^{-3}$\\
\hline
\end{tabular}
\end{table}

\section{Theoretical Details of Decay Amplitudes}

\begin{center}
{\bf A. Within the Standard Model}
\end{center}

\noindent
The decay amplitudes for the processes of Fig.~1, with $M_i$ listed in Table 1,
%
\be
B^-(p_B)\to M_i(p_i,\l_i)+W^*(q)\to
 \pi^+(p_+)+\pi^-(p_-)+l^-(p_l,\l_l)+\bar{\nu}(p_\nu)
\ee
are expressed as
\be
{\cal A}^{\lambda_l}&=&-\frac{G_F}{\sqrt{2}}\sum_{i}\sum_{\lambda_i}
     V_i c_i \langle l^-(p_l,\lambda_l)\bar{\nu}(p_\nu)|j^{\mu\dagger}|0\rangle 
             \langle M_i(p_i,\lambda_i)|J_{i\mu}|B^-(p_B)\rangle \nonumber\\
         & &\times \Pi_i(s_{_M})\langle \pi^+(p_+)\pi^-(p_-)\|M_i(p_i,\lambda_i)\rangle ,
\ee
where $\l_i=0$ for spin $0$ states ($f_0$ and $D$),
$\l_i=\pm 1,0$ for spin $1$ states ($\rho$),
$\l_i=\pm 2,\pm 1,0$ for spin $2$ states ($f_2$),
and $\l_l$ is the lepton helicity, $\pm\frac{1}{2}$.

\begin{figure}[ht]
\hbox to\textwidth{\hss\epsfig{file=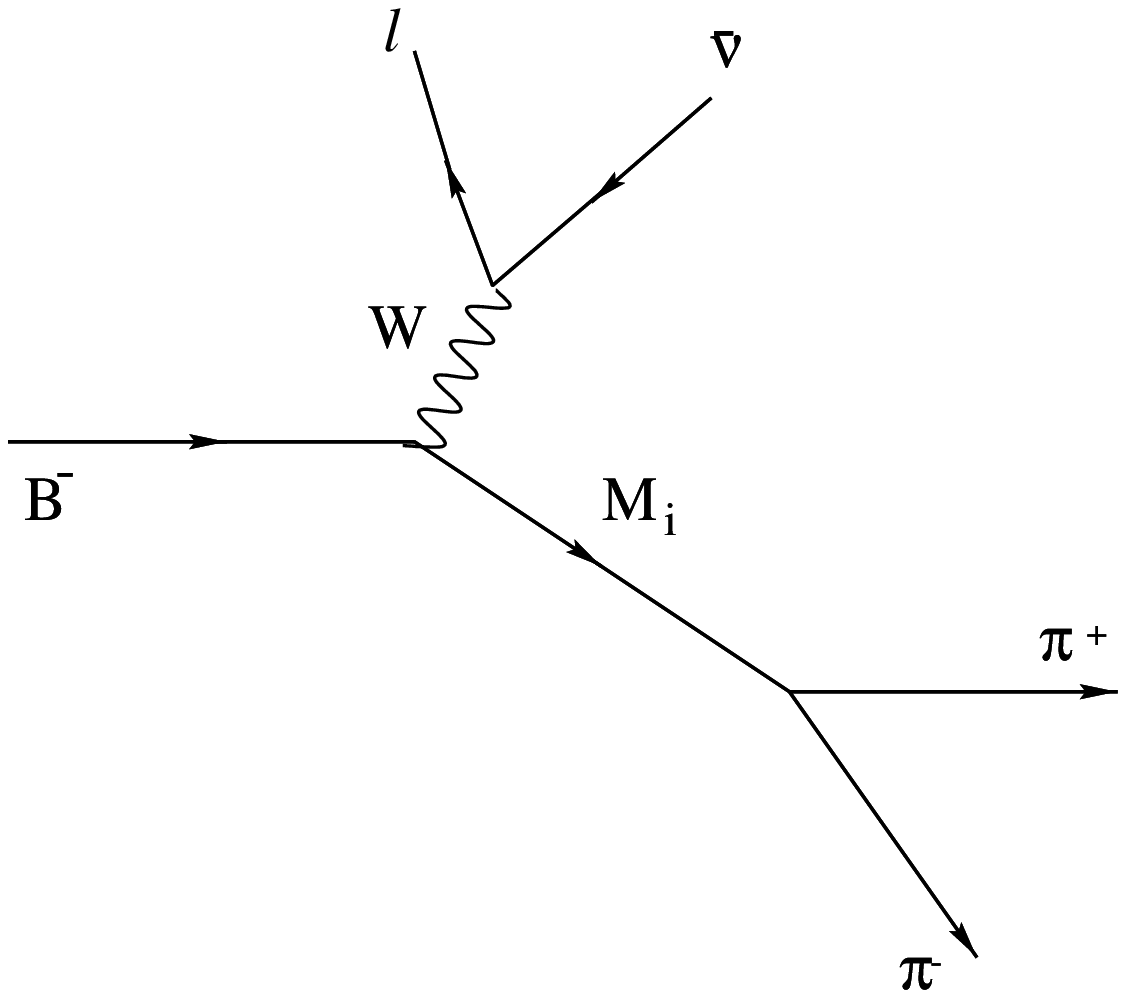,height=9cm,width=8cm,angle=-90}\hss}
\caption{Diagrams for $B\to M_iW^*\to \pi^+\pi^- l^-\bar{\nu}_l$ decays.}
\label{fig:diagram}
\end{figure}

The leptonic current is
\be
j^\mu = \bar{\psi}_\nu \gamma^\mu(1-\gamma_5)\psi_l,
\ee
and for the hadronic currents we have
\be
J_i^\mu = \bar{\psi}_u \gamma^\mu(1-\gamma_5)\psi_b, \;\; V_i=V_{ub},
         \;\; c_i=\frac{1}{\sqrt{2}}\;\;\;
         {\rm for}\;\; {\rm label}~~i=0^{^{(\prime)}},1^{^{(\prime)}},2;
\ee
and
\be
J_3^\mu = \bar{\psi}_c \gamma^\mu(1-\gamma_5)\psi_b, \;\; V_3=V_{cb},
         \;\; c_3=1\;\;\; {\rm for}\;\; {\rm label}~~i=3,
\ee
where $c_i$ stands for the isospin factor especially due to $u\bar{u}$-mesons.
We assume that the resonance contributions of the intermediate states
can be treated by the Breit-Wigner form, which is written
in the narrow width approximation as
\be
\Pi_i (s_{_M})=\frac{\sqrt{m_i\Gamma_i/\pi}}{s_{_M}-m_i^2+im_i\Gamma_i},
\ee
where $s_{_M}=(p_+ + p_-)^2$ and the $m_i$'s and $\Gamma_i$'s are the masses
and widths of the resonances respectively (See Table 1).
For the decay parts of the resonances we use \cite{cancel}
\be
\langle \pi^+(p_+)\pi^-(p_-)\|M_i(p_i,\lambda_i)\rangle 
=\sqrt{{\cal BR}_i}Y^{\lambda_i}_{\lambda_i max} (\theta^*,\phi^*),
\ee
where $Y^m_l(\theta,\phi)$ are the $J=l$ spherical harmonics listed in Appendix B,
and the angles $\theta^*$
and $\phi^*$ are those of the final state $\pi^-$ specified in the $M_i$
rest frame (See Fig. 2c).
The couplings of $M_i$ to $\pi\pi$ are effectively taken into account by
the branching fractions, ${\cal BR}_i(M_i\to \pi^+\pi^-)$.

\begin{figure}[tb]
\hbox to\textwidth{\hss\epsfig{file=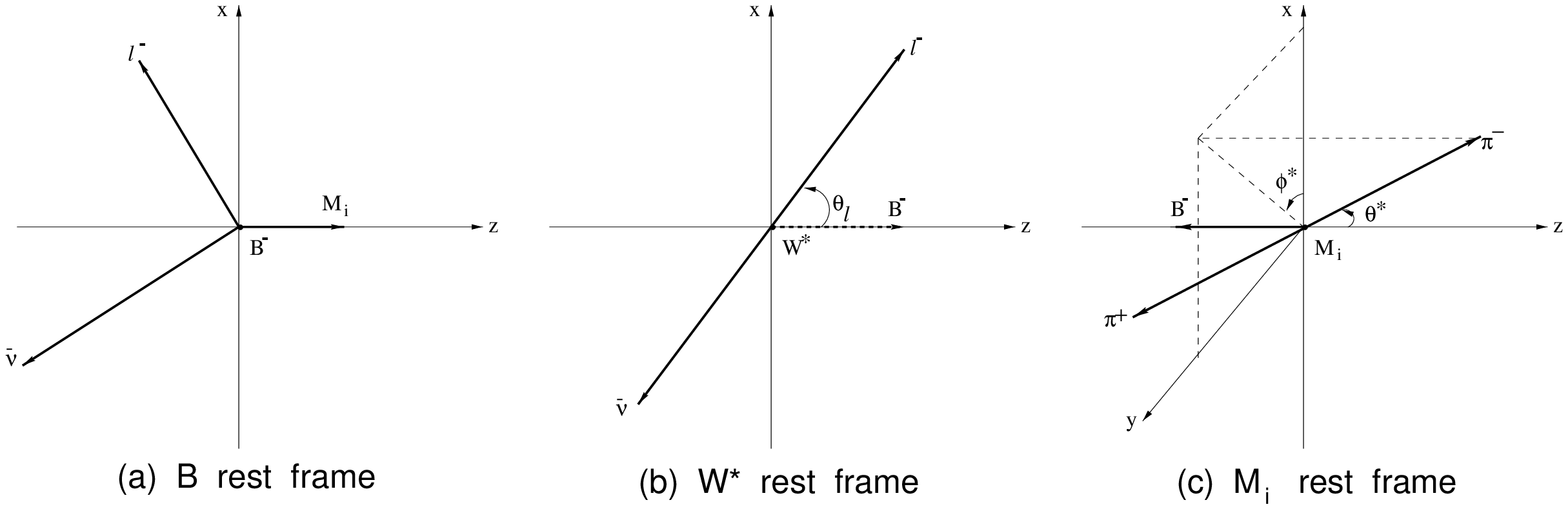,height=15cm,width=5.5cm,angle=-90}\hss}
\caption{The decay $B\to M_iW^*\to (\pi^+\pi^-)(l\bar{\nu})$ viewed from the
(a) $B^-$, (b) $W^*$ and (c) $M_i$ rest frames.}
\label{fig:frame}
\end{figure}

In order to obtain the full helicity amplitude of the $B\to \pi\pi l\nu$ decay,
we first consider the amplitude of $B\to M_i l \bar{\nu}_l$ \cite{bl3},
denoted as ${\cal M}^{\l_l}_{\l_i}$:
\be
{\cal M}^{\l_l}_{\l_i}=-\frac{G_F}{\sqrt{2}}
     V_i c_i \langle l^-(p_l,\lambda_l)\bar{\nu}(p_\nu)|j^{\mu\dagger}|0\rangle 
             \langle M_i(p_i,\lambda_i)|J_{i\mu}|B^-(p_B)\rangle .
\ee
We express the matrix elements
${\cal M}^{\l_l}_{\l_i}$ into the following form:
\be
{\cal M}^{\l_l}_{\l_i}=\frac{G_F}{\sqrt{2}}V_i c_i
       \sum_{\l_W}\eta_{\l_W}L^{\l_l}_{\l_W}H^{\l_i}_{\l_W},
\label{smamp}
\ee
where for the decays $B \to M_i W^*$ and $W^* \to l \bar{\nu}$, respectively,
\be
H^{\l_i}_{\l_W}&=&\e^*_{W\mu}\langle M_i(p_i,\l_i)|J^\mu_i|B^-(p_B)\rangle ,\nonumber\\
L^{\l_l}_{\l_W}&=&\e_{W\mu}\langle l^-(p_l,\l_l)\bar{\nu}(p_\nu)|j^{\mu\dagger}|0\rangle ,
\ee
in terms of the polarization vectors $\e_W\equiv \e(q,\l_W)$ of the virtual $W$.
These $\e_W$'s satisfy the relation 
\be
-g^{\mu\nu}=\sum_{\l_W}\eta_{\l_W}\e^\mu_W \e^{*\nu}_W,
\label{metric}
\ee
where the summation is over the helicities  $\l_W =\pm 1,0,s$ of the virtual $W$, 
with the metric $\eta_\pm=\eta_0=-\eta_s=1$.

We evaluate the leptonic amplitude $L^{\l_l}_{\l_W}$ in the rest frame of
the virtual $W$ (See Fig. 2b) with the $z$-axis chosen along the $M_i$ 
direction, and the $x$--$z$ plane chosen as the virtual $W$ decay plane,
with $(p_l)_x>0$.
Using the 2-component spinor technique \cite{hagiwara} and
polarization vectors given in  Appendix B, we find
\be
&&L^-_\pm=2\sqrt{q^2}vd_\pm,\; L^-_0=-2\sqrt{q^2}vd_0,\; \hspace{1.1cm} L^-_s=0,\nonumber\\
&&L^+_\pm=\pm 2m_lvd_0,\; L^+_0=\sqrt{2}m_lv(d_+-d_-),\; L^+_s=-2m_lv,
\label{Lpm}
\ee
where
\be
v=\sqrt{1-\frac{m_l^2}{q^2}},\; d_\pm=\frac{1\pm\cos\theta_l}{\sqrt{2}},\;
~~{\rm and}~~d_0=\sin\theta_l.
\ee
Here we show only the sign of $\l_l$ as a superscript on $L$.
Note that the $L^+$ amplitudes are proportional
to the lepton mass $m_l$, and the scalar amplitude $L^-_s$ vanishes due to
angular momentum conservation.

For $B\to M_i$ transitions through the weak charged current 
\be
J_i^\mu=V^\mu_i-A^\mu_i,
\ee
the most general forms of matrix elements are
\be
{\rm for}&f_0(0^+)&{\rm states}:\nonumber\\
&&\langle f_0(p_i)|V_\mu|B(p_B)\rangle =0,\nonumber\\
&&\langle f_0(p_i)|A_\mu|B(p_B)\rangle 
 =u_+(q^2)(p_B+p_i)_\mu+u_-(q^2)(p_B-p_i)_\mu;\nonumber\\
{\rm for}&\rho(1^-)&{\rm states}:\nonumber\\
&&\langle \rho(p_i,\e_1)|V_\mu|B(p_B)\rangle 
 =ig(q^2)\e_{\mu\nu\rho\sigma}\e_1^{*\nu}(p_B+p_i)^\rho
                           (p_B-p_i)^\sigma,\nonumber\\
&&\langle \rho(p_i,\e_1)|A_\mu|B(p_B)\rangle 
 =f(q^2)\e^*_{1\mu}+a_+(q^2)(\e_1^*\cdot p_B)(p_B+p_i)_\mu\nonumber\\
    &&\hspace{4.5cm} +a_-(q^2)(\e_1^*\cdot p_B)(p_B-p_i)_\mu;\nonumber\\
{\rm for}&f_2(2^+)&{\rm states}:\nonumber\\
&&\langle f_2(p_i,\e_2)|V_\mu|B(p_B)\rangle 
 =ih(q^2)\e_{\mu\nu\l\rho}\e_2^{*\nu\alpha}p_{B\alpha}
                  (p_B+p_i)^\l(p_B-p_i)^\rho,\nonumber\\
&&\langle f_2(p_i,\e_2)|A_\mu|B(p_B)\rangle 
 =k(q^2)\e^*_{2\mu\nu}p_B^\nu +b_+(q^2)(\e^*_{2\alpha\beta}
                   p_B^\alpha p_B^\beta)(p_B+p_i)_\mu\nonumber\\
 &&\hspace{4.5cm} +b_-(q^2)(\e^*_{2\alpha\beta}p_B^\alpha p_B^\beta)(p_B-p_i)_\mu;\nonumber\\
{\rm for}&D(0^-)&{\rm states}:\nonumber\\
&&\langle D(p_i)|V_\mu|B(p_B)\rangle =f_+(q^2)(p_B+p_i)_\mu+f_-(q^2)(p_B-p_i)_\mu,\nonumber\\
&&\langle D(p_i)|A_\mu|B(p_B)\rangle =0,
\label{BMamp}
\ee
where $\e_1$ and $\e_2$ are the polarization vectors of the spin 1 
and spin 2 states, respectively.
Using the above expressions and the polarization vectors given in Appendix B, 
we find non-zero $B\to M_iW^*$ amplitudes are
\be
{\rm for}&i=0,&\;\; H^0_{\l_W}\equiv S^0_{\l_W},\nonumber\\
&&S^0_0=-u_+(q^2)\frac{\sqrt{Q_+Q_-}}{\sqrt{q^2}},\;
  S^0_s=-\left(u_+(q^2)\frac{(m_B^2-s_{_M})}{\sqrt{q^2}}+u_-(q^2)\sqrt{q^2}\right),
\ee
\be
{\rm for}&i=1^{^{(\prime)}},&H^{\l_1}_{\l_W}\equiv V^{\l_1}_{\l_W},\nonumber\\
&&V^0_0=-\frac{1}{2\sqrt{s_{_M}q^2}}
 \left[f(q^2)(m_B^2-s_{_M}-q^2)+a_+(q^2)Q_+Q_-\right],\nonumber\\
&&V^{\pm 1}_{\pm 1}=f(q^2)\mp g(q^2)\sqrt{Q_+Q_-},\nonumber\\
&&V^0_s=-\frac{\sqrt{Q_+Q_-}}
 {2\sqrt{s_{_M}q^2}}\left[f(q^2)+a_+(q^2)(m_B^2-s_{_M})+a_-(q^2)q^2\right],
\ee
\be
{\rm for}&i=2,&H^{\l_2}_{\l_W}\equiv T^{\l_2}_{\l_W},\nonumber\\
&&T^0_0=-\frac{1}{2\sqrt{6}}\frac{\sqrt{Q_+Q_-}}{s_{_M}\sqrt{q^2}}
         \left[k(q^2)(m_B^2-s_{_M}-q^2)+b_+(q^2)Q_+Q_-\right],\nonumber\\
&&T^{\pm 1}_{\pm 1}=\frac{1}{2\sqrt{2}}\sqrt{\frac{Q_+Q_-}{s_{_M}}}
          [k(q^2)\mp h(q^2)\sqrt{Q_+Q_-}],\nonumber\\
&&T^0_s=-\frac{1}{2\sqrt{6}}\frac{Q_+Q_-}{s_{_M}\sqrt{q^2}}
         \left[k(q^2)+b_+(q^2)(m_B^2-s_{_M})+b_-(q^2)q^2\right],
\ee
\be
{\rm for}&i=3,&H^0_{\l_W}\equiv P^0_{\l_W},\nonumber\\
&&P^0_0=f_+(q^2)\frac{\sqrt{Q_+Q_-}}{\sqrt{q^2}},\;
  P^0_s=f_+(q^2)\frac{(m_B^2-s_{_M})}{\sqrt{q^2}}+f_-(q^2)\sqrt{q^2},
\ee
where 
\be
Q_\pm=(m_B\pm\sqrt{s_{_M}})^2-q^2.
\label{Qpm}
\ee
Here 
\be
Q_+Q_-=\l(m_B^2,s_{_M},q^2)
\ee
gives the triangle function $\l(a,b,c)=a^2+b^2+c^2-2(ab+bc+ca)$.
Combining all the formulae, we can write the full helicity amplitudes of
$B^-\to \pi^+\pi^- l^-\bar{\nu}$ decays as
\be
{\cal A}^{\lambda_l}&=&V_{ub}\frac{G_F}{\sqrt{2}}\bigg[
      \sum_{\l=0,s} \eta_\l L^{\l_l}_\l (\Pi_{f_0} S^0_\l Y^0_0 +
      \xi \Pi_D P^0_\l \tilde{Y}^0_0+\Pi_\rho V^0_\l Y^0_1
        +\Pi_{f_2} T^0_\l Y^0_2)\nonumber\\
      &&+\sum_{\l=\pm 1} L^{\l_l}_\l (\Pi_\rho V^\l_\l Y^\l_1
        +\Pi_{f_2} T^\l_\l Y^\l_2)\bigg],
\label{amp}
\ee
where
\be
\Pi_{f_0}&\equiv&\frac{1}{\sqrt{2}}(\sqrt{{\cal BR}_0}\Pi_0
           +\sqrt{{\cal BR}_{0^\prime}}\Pi_{0^\prime})\nonumber\\
\Pi_{\rho}&\equiv&\frac{1}{\sqrt{2}}(\sqrt{{\cal BR}_1}\Pi_1
           +\sqrt{{\cal BR}_{1^\prime}}\Pi_{1^\prime})\nonumber\\
\Pi_{f_2}&\equiv&\frac{1}{\sqrt{2}}\sqrt{{\cal BR}_2}\Pi_2\nonumber\\
\Pi_D&\equiv&\sqrt{{\cal BR}_3}\Pi_3,
\ee
and 
\be
\xi=\frac{V_{cb}}{V_{ub}}.
\label{xi}
\ee
Note that we use $\tilde{Y}^0_0$ for the pseudo scalar
meson $D$, which is actually the same quantity as $Y^0_0=1/\sqrt{4\pi}$ 
for the scalar meson $f_0$
except that it changes sign under the parity transformation.           
Concerning the parametrization of $\xi$, 
other CKM factors, such as $V_{cd}^*$ from $D^0\to \pi^+\pi^-$ decay, 
are already included in its branching fraction calculation.
And because we use implicitly Wolfenstein parametrization \cite{wolfenstein} 
for CKM matrix, in which the complex phases are approximately in the elements 
$V_{td}$ and $V_{ub}$ only,
the imaginary part of $\xi$ here comes only from the element $V_{ub}$.

The differential partial width of interest can be expressed as
\be
d\Gamma(B^- \to \pi^+ \pi^- l^- \bar{\nu}_l)
 =\frac{1}{2m_B}\sum_{\l_l}|{\cal A}^{\l_l}|^2
       \frac{(q^2-m_l^2)\sqrt{Q_+Q_-}}{256\pi^3m_B^2q^2}d\Phi_4,       
\ee
where the 4 body phase space $d \Phi_4$ is
\be
d \Phi_4 \equiv  ds_{_M} \cdot dq^2 \cdot d\cos\theta^* \cdot 
                 d\cos\theta_l \cdot d\phi^*.
\ee
Kinematically allowed regions of the variables are
\be
&&4m_\pi^2\;<\;s_{_M}\;<\;(m_B-m_l)^2,\nonumber\\
&&m_l^2\;<\;q^2\;<\;(m_B-\sqrt{s_{_M}})^2,\nonumber\\
&&-1\;<\;\cos\theta^*,\;\cos\theta_l\;<\;1,\nonumber\\
&&0\;<\;\phi^*\;<\;2\pi .
\ee

Since the initial $B^-$ system is not CP self-conjugate, any genuine
CP-odd observable can be constructed only by considering both the $B^-$
decay and its charge-conjugated $B^+$ decay, and by identifying the CP
relations of their kinematic distributions.
Before constructing possible CP-odd asymmetries explicitly, we calculate
the decay amplitudes for the charge-conjugated process
$B^+\to \pi^+ \pi^- l^+ \nu_l$. 
For the charge-conjugated $B^+$ decays,
the amplitudes can be written as
\be
{\bar{\cal A}}^{\l_l}&=&-\frac{G_F}{\sqrt{2}}\sum_{i}\sum_{\l_i}
   V_i^* c_i \langle l^+(p_l,\lambda_l)\nu(p_\nu)|j^{\mu}|0\rangle 
    \langle \overline{M}_i(p_i,\lambda_i)|J^\dagger_{i\mu}|B^+(p_B)\rangle \nonumber\\
  & &\times \Pi_i(s_{_M})\langle \pi^+(p_+)\pi^-(p_-)\|\overline{M}_i(p_i,\lambda_i)\rangle .
\ee
The leptonic amplitudes $\bar{L}^{\l_l}_{\l_W}$ are
\be
&&\bar{L}^+_\pm=-2\sqrt{q^2}vd_\mp,\; \bar{L}^+_0
   =-2\sqrt{q^2}vd_0,\; \hspace{1.1cm} \bar{L}^+_s=0,\nonumber\\
&&\bar{L}^-_\pm=\pm 2m_lvd_0,\;\;\;\; \bar{L}^-_0=\sqrt{2}m_lv(d_+-d_-),\; \bar{L}^-_s=-2m_lv.
\ee
And the transition amplitudes $\overline{H}^{\l_i}_{\l_W}$ for $B^+\to \overline{M}_iW^*$
are given by a simple modification of the 
amplitudes $H^{\l_i}_{\l_W}$ of the $B^-$ decays:
\be
\overline{H}^{\l_i}_{\l_W}=H^{\l_i}_{\l_W}\{g\to -g,\;h\to -h,\;f_\pm\to-f_\pm\}.
\ee
Then, we find the full amplitude for $B^+\to \pi^+\pi^-l^+\nu$:
\be
{\bar{\cal A}}^{\lambda_l}&=&V_{ub}^*\frac{G_F}{\sqrt{2}}\bigg[
      \sum_{\l=0,s} \eta_\l\bar{L}^{\l_l}_\l (\Pi_{f_0} \overline{S}^0_\l Y^0_0 +
      \xi^* \Pi_D \overline{P}^0_\l \tilde{Y}^0_0+\Pi_\rho \overline{V}^0_\l Y^0_1
        +\Pi_{f_2} \overline{T}^0_\l Y^0_2)\nonumber\\
     && +\sum_{\l=\pm 1} \bar{L}^{\l_l}_\l (\Pi_\rho \overline{V}^\l_\l Y^\l_1
        +\Pi_{f_2} \overline{T}^\l_\l Y^\l_2)\bigg],
\label{ampbar}
\ee
where
\be
&&\overline{S}^0_{\l_W}=S^0_{\l_W},\nonumber\\
&&\overline{P}^0_{\l_W}=-P^0_{\l_W},\nonumber\\
&&\overline{V}^0_{0,s}=V^0_{0,s},\; \overline{V}^{\pm 1}_{\pm 1}=V^{\mp 1}_{\mp 1},\nonumber\\
&&\overline{T}^0_{0,s}=T^0_{0,s},\; \overline{T}^{\pm 1}_{\pm 1}=T^{\mp 1}_{\mp 1}.
\ee

It is easy to see that
if $V_{ub}$ and $V_{cb}$ are real, the amplitude (\ref{amp}) of the $B^-$ decay and
(\ref{ampbar}) of the $B^+$ decay satisfy the CP relation:
\be
{\cal A}^{\pm}(\theta^*,\phi^*,\theta_l)
   =\eta_{CP}{\bar{\cal A}}^{\mp}(\theta^*,-\phi^*,\theta_l;
               \tilde{Y}^0_0\to-\tilde{Y}^0_0),
\label{cprel}
\ee
where $\theta^*$ and $\phi^*$ in $\bar{\cal A}^{\l_l}$ are the angles of
the final state $\pi^+$, while those in ${\cal A}^{\l_l}$ are for $\pi^-$.  
Then, with a complex weak phase $\xi$, 
$d\Gamma/d\Phi_4$ can be decomposed into a CP-even part ${\cal S}$ and
a CP-odd part ${\cal D}$:
\be
\frac{d\Gamma}{d\Phi_4}=\frac{1}{2}({\cal S}+{\cal D}).
\ee
The CP-even part ${\cal S}$ and the CP-odd part ${\cal D}$ can be easily
identified by making use of the CP relation (\ref{cprel}) between $B^-$ and 
$B^+$ decay amplitudes, and they are expressed as
\be
{\cal S} =\frac{d(\Gamma+\overline{\Gamma})}{d\Phi_4},\quad
{\cal D} =\frac{d(\Gamma-\overline{\Gamma})}{d\Phi_4},
\ee
where $\Gamma$ and $\overline{\Gamma}$ are the decay rates for $B^-$ and $B^+$, 
respectively, and 
we have used the same kinematic variables $\{s_{_M},q^2,\theta^*,\theta_l\}$
for the $d\overline{\Gamma}/d\Phi_4$ except for the replacements of 
${\phi^*} \to -\phi^*$ and $\tilde{Y}^0_0 \to -\tilde{Y}^0_0$, 
as shown in Eq. (\ref{cprel}).
The CP-even ${\cal S}$ term and the CP-odd ${\cal D}$ term can be obtained from
$B^\mp$ decay probabilities, and their explicit form is listed in Appendix C.
Note that the CP-odd term is proportional to the imaginary part 
of the parameter $\xi$ in Eq.~(\ref{xi}).

Before we go further on to the beyond the SM analyses, 
we note that in addition to the resonant tree diagram contributions
there are other SM contributions through
annihilation diagrams and electroweak penguin diagrams, 
which are relevant for nonresonant case.
As written in Section 1, we consider only resonant contributions by assuming
nonresonant contributions can be separated through data analyses.

\begin{center}
{\bf B. With complex scalar couplings}
\end{center} 

Next we consider CP violation effects in extensions of the SM, where we extend
the virtual $W$-exchange part in Fig.~1 by including an additional scalar
interaction with complex couplings.
First we describe the formalism in a model independent way, but later
we consider specific models such as multi-Higgs doublet models and
scalar-leptoquark models.
In this case CP-violating phases can be generated through the interference
between $W$-exchange diagrams and scalar exchange diagrams with complex couplings.

The decay amplitudes for $B^- \to \pi^+ \pi^- l^- \bar{\nu}_l$ are expressed as
\be
{\cal A}^{\lambda_l}&=&-V_{ub}\frac{G_F}{\sqrt{2}}\frac{1}{\sqrt{2}}\sum_{i}\sum_{\lambda_i}
     \bigg[\langle l^-(p_l,\lambda_l)\bar{\nu}(p_\nu)|j^{\mu\dagger}|0\rangle 
             \langle M_i(p_i,\lambda_i)|J_{\mu}|B^-(p_B)\rangle \nonumber\\
        & &+\zeta\langle l^-(p_l,\lambda_l)\bar{\nu}(p_\nu)|j_s^{\dagger}|0\rangle 
             \langle M_i(p_i,\lambda_i)|J_s|B^-(p_B)\rangle \bigg]\nonumber\\
         & &\times \Pi_i(s_{_M})\langle \pi^+(p_+)\pi^-(p_-)\|M_i(p_i,\lambda_i)\rangle ,
\ee
where
\be
&&j^\mu=\bar{\psi}_\nu\gamma^\mu(1-\gamma_5)\psi_l,\nonumber\\
&&J^\mu=\bar{\psi}_u\gamma^\mu(1-\gamma_5)\psi_b\equiv V^\mu-A^\mu,
\ee
and their corresponding scalar currents are
\be
j_s=\bar{\psi}_\nu(1-\gamma_5)\psi_l,\;\;
J_s=\bar{\psi}_u(1-\gamma_5)\psi_b,
\ee
the additional factor $1/\sqrt{2}$ comes from the isospin factor as mentioned earlier.
Here the parameter $\zeta$, which parameterizes contributions from physics beyond
the SM, is in general a complex number.
And as explained earlier, 
in order to exclude any possible CP violation effects induced within the SM, 
we only include the lowest three states $\rho(770)$, $f_0(980)$
and $f_2(1270)$ as intermediate states.
By using the Dirac equation for the leptonic current, $q_\mu j^\mu=m_l j_s$,
the amplitude can be written as
\be
{\cal A}^{\lambda_l}&=&-V_{ub}\frac{G_F}{2}\sum_{i}\sum_{\lambda_i}
     \langle l^-(p_l,\lambda_l)\bar{\nu}(p_\nu)|j^{\mu\dagger}|0\rangle 
             \langle M_i(p_i,\lambda_i)|\Omega_\mu|B^-(p_B)\rangle \nonumber\\
         & &\times \Pi_i(s_{_M})\langle \pi^+(p_+)\pi^-(p_-)\|M_i(p_i,\lambda_i)\rangle ,
\ee
where the effective hadronic current $\Omega_\mu$ is defined as
\be
\Omega_\mu\equiv J_\mu + \zeta \frac{q_\mu}{m_l} J_s.
\ee
In this case the amplitudes ${\cal M}^{\l_l}_{\l_i}$ of $B\to M_i l\bar{\nu}$
have the same form as the previous SM case (\ref{smamp}) except for the 
modification in the hadronic current part due to the additional scalar current:
\be
{\cal M}^{\l_l}_{\l_i}=\frac{G_F}{2}V_{ub}
       \sum_{\l_W}\eta_{\l_W}L^{\l_l}_{\l_W}{\cal H}^{\l_i}_{\l_W},
\ee
where ${\cal H}^{\l_i}_{\l_W}$ stands for the hadronic amplitudes modified 
by the scalar current $J_s$.
Using the equation of motion for $u$ and $b$ quarks, we get within the on-shell
approximation
\be
J_s = (p_b^\mu-p_u^\mu) \left[\frac{V_\mu}{m_b-m_u}+\frac{A_\mu}{m_b+m_u}\right] .
\ee
Later for numerical calculations,
we use the approximation, 
$(p_b^\mu-p_u^\mu) \approx (p_B^\mu-p_{M_i}^\mu) \equiv q^\mu$,
which is assumed in quark model calculations of form factors \cite{ISGW}.
After explicit calculation, we find that the additional scalar current modifies
only the scalar component of ${\cal H}^{\l_i}_{\l_W}$: {\it i.e.}
\be
{\cal H}^0_s&=&(1-\zeta^\prime) H^0_s,\nonumber\\
{\rm and}~~~{\cal H}^{\l_i}_{\l_W}&=&H^{\l_i}_{\l_W}
        \;\;{\rm for}\;\; \l_W = 0,~\pm 1,
\ee
where
\be
\zeta^\prime=\frac{q^2}{m_l(m_b+m_u)}\zeta .
\label{zetap}
\ee
In this case, 
$d\Gamma/d\Phi_4$ also can be decomposed into a CP-even part ${\cal S}$ and
a CP-odd part ${\cal D}$:
\be
\frac{d\Gamma}{d\Phi_4}=\frac{1}{2}({\cal S}+{\cal D}).
\ee
Their explicit form is listed in Appendix C.
Note that the CP-odd term is proportional to the imaginary part 
of the parameter $\zeta$ and the lepton mass $m_l$.
Therefore, we have to consider massive leptonic ($\mu$ or $\tau$) decays.

As specific extensions of the SM,
we consider four types of scalar-exchange models which preserve the 
symmetries of the SM \cite{scalar}:
One of them is the multi-Higgs-doublet (MHD) model \cite{Grossman} and
the other three models are scalar-leptoquark (SLQ) models \cite{wyler,randall}.
The authors of Ref.~\cite{tau} investigated CP violations in $\tau$ decay processes
with these extended models. We follow their description and 
make it to be appropriate for our analysis.

In the MHD model CP violation can arise in the charged Higgs sector
with more than two Higgs doublets \cite{Wein} and when not all the charged 
scalars are degenerate. As in most previous phenomenological analyses,
we  assume that all but the lightest of the charged
scalars effectively decouple from fermions. 
The effective Lagrangian of the MHD model contributing 
to the decay $B\rightarrow \pi\pi l\bar{\nu}_l$ 
is then given at energies considerably low compared to $M_H$  by 
\be
{\cal L}_{_{MHD}}=2\sqrt{2}G_FV_{ub}\frac{m_l}{M_H^2}\Big[m_b XZ^*(\bar{u}_L b_R)
            +m_u YZ^*(\bar{u}_R b_L)\Big](\bar{l}_R \nu_L),
\ee
where $X$, $Y$ and $Z$ are complex coupling constants which can be
expressed in terms of the charged Higgs mixing matrix elements.
{}From the effective Lagrangian, we obtain for the MHD CP-violation parameter
${\rm Im}(\zeta_{_{MHD}})$, 
\be                                            
\Im(\zeta_{_{MHD}})=\frac{m_l m_b}{M_H^2}\Big\{\Im(XZ^*)-(\frac{m_u}{m_b})\Im(YZ^*)\Big\}.
\label{etaMHD}
\ee
The constraints on the CP-violation parameter (\ref{etaMHD}) depend upon 
the values chosen for the $u$ and $b$ quark masses.
In the present work, we use (See Appendix A)
\be
m_u=0.33\;{\rm GeV},\qquad m_b=5.12\;{\rm GeV}.
\ee
In the MHD model the strongest constraint \cite{Grossman} on $\Im(XZ^*)$
comes from the measurement of the branching ratio 
${\cal B}(b\to X\tau \nu_\tau)$, which actually gives a constraint on $|XZ|$. 
For $M_H<440$ GeV, the bound on $\Im(XZ^*)$ is given by
\be
\Im(XZ^*)\;<\;|XZ|\;<0.23M^2_H \;{\rm GeV}^{-2}.
\ee
On the other hand, the bound on $\Im(YZ^*)$ is mainly
given by $K^+\to \pi^+\nu\bar{\nu}$. The present bound \cite{Grossman} is
\be
\Im(YZ^*)\;<\;|YZ|\;<110.
\ee
Combining the above bounds, we obtain
the following bounds on $\Im(\zeta_{_{MHD}})$ as
\be
&&|\Im(\zeta_{_{MHD}})|\;<\; 2.06\;\qquad\qquad {\rm for}\;\tau\;{\rm family},\nonumber\\
&&|\Im(\zeta_{_{MHD}})|\;<\; 0.12\;\qquad\qquad {\rm for}\;\mu\;{\rm family}.
\label{Hmodel}
\ee

On the other hand, the effective Lagrangians for the three SLQ models \cite{scalar,wyler}
contributing to the decay $B\to \pi\pi l\nu$ are written in the form, 
after a few Fierz rearrangements:
\be
{\cal L}^{^{I}}_{_{SLQ}}&=&-\frac{x_{3j}x^{\prime *}_{1j}}{2M^2_{\phi_1}}\left[
             (\bar{b}_Lu_R)(\bar{\nu}_{lL}l_R)
            +\frac{1}{4}(\bar{b}_L\sigma^{\mu\nu}u_R)
             (\bar{\nu}_{lL}\sigma_{\mu\nu}l_R)\right]+h.c.,\nonumber\\
{\cal L}^{^{II}}_{_{SLQ}}&=&-\frac{y_{3j}y^{\prime *}_{1j}}{2M^2_{\phi_2}}\left[
             (\bar{b}_Lu_R)(\bar{l}^c_R\nu^c_{l L})
            +\frac{1}{4}(\bar{b}_L\sigma^{\mu\nu}u_R)
          (\bar{l}^c_R\sigma_{\mu\nu}\nu^c_{l L})\right]\nonumber\\
          &&+\frac{y_{3j}y^*_{1j}}{2M^2_{\phi_2}}(\bar{b}_L\gamma_\mu u_L)
           (\bar{l}^c_L\gamma^\mu\nu^c_{l L})+h.c.,\nonumber\\
{\cal L}^{^{III}}_{_{SLQ}}&=&-\frac{z_{3j}z^*_{1j}}{2M^2_{\phi_3}}(\bar{b}_L\gamma_\mu u_L)
           (\bar{l}^c_L\gamma^\mu\nu^c_{l L})+h.c.\;,
\ee
where $j=2,3$ for $l=\mu,\tau$, respectively 
and the coupling constants $x^{(\prime)}_{ij}$, $y^{(\prime)}_{ij}$ and
$z_{ij}$ are in general complex so that CP is violated
in the scalar-fermion Yukawa interaction terms. The superscript $c$ in the
Lagrangians ${\cal L}^{^{II}}_{_{SLQ}}$ and ${\cal L}^{^{III}}_{_{SLQ}}$ denotes 
charge conjugation, {\it i.e.} $\psi^c_{R,L}=i\gamma^0\gamma^2\bar{\psi}^T_{R,L}$ 
in the chiral representation. 
Then we find that the size of the SLQ model
CP-violation effects is dictated by the CP-odd parameters
\begin{eqnarray}
&&{\rm Im}(\zeta^{^I}_{_{SLQ}})=-\frac{{\rm Im}[x_{3j}
      x^{\prime *}_{1j}]}{4\sqrt{2}G_FV_{ub}M^2_{\phi_1}}\,,\nonumber\\
&&{\rm Im}(\zeta^{^{II}}_{_{SLQ}})= -\frac{{\rm Im}[y_{3j}y^{\prime *}_{1j}]}
   {4\sqrt{2}G_FV_{ub}M^2_{\phi_2}}\,,\nonumber\\
&&{\rm Im}(\zeta^{^{III}}_{_{SLQ}})= 0\,.
\label{LQp}
\end{eqnarray}
Although there are at present no direct 
constraints on the SLQ model CP-odd parameters in (\ref{LQp}), 
a rough constraint to the parameters can be provided by the assumption 
\cite{Davidson} that $|x^\prime_{1j}|\sim |x_{1j}|$ and 
$|y^\prime_{1j}|\sim |y_{1j}|$, that is to say, the leptoquark couplings   
to quarks and leptons belonging to the same generation are of a similar
size; then the experimental upper bounds 
from $B\bar{B}$ mixing for $\tau$ family, $B\to \mu\bar{\mu}X$ decay 
for $\mu$ in Model I, and $B\to l\nu X$ for $\tau$
together with the $V_{ub}$ measurement
for $\mu$ in Model II yield \cite{Davidson}
\begin{eqnarray}
&&|{\rm Im}(\zeta^{^I}_{_{SLQ}})|< 2.76,\quad 
|{\rm Im}(\zeta^{^{II}}_{_{SLQ}})|< 18.4 \qquad\; {\rm for}\;\tau\;{\rm family},\nonumber\\
&&|{\rm Im}(\zeta^{^I}_{_{SLQ}})|< 0.37,\quad 
|{\rm Im}(\zeta^{^{II}}_{_{SLQ}})|< 1.84 \qquad\; {\rm for}\;\mu\;{\rm family}.
\label{Models}
\end{eqnarray}
Based on the constraints (\ref{Hmodel}) and (\ref{Models}) to the CP-odd
parameters, we quantitatively estimate the number of
$B^-\to \pi^+\pi^- l^-\bar{\nu}_l$ decays to detect CP violation 
for the maximally-allowed values of the CP-odd parameters.

\section{Observable CP Asymmetries}

\noindent
An easily constructed CP-odd asymmetry is the rate asymmetry
\be
A\equiv \frac{\Gamma-\overline{\Gamma}}{\Gamma+\overline{\Gamma}},
\ee
which has been used as a 
probe of CP violation in Higgs and top quark sectors  \cite{rateasym}.  
Here $\Gamma$ and $\overline{\Gamma}$ are the decay rates for $B^-$ and $B^+$, 
respectively.
The statistical significance of the asymmetry can be computed as
\be
N_{SD}=\frac{N_- -N_+}{\sqrt{N_-+N_+}} =\frac{N_- -N_+}{\sqrt{N \cdot Br}},
\ee
where $N_{SD}$ is the number of standard deviations,
$N_\pm$ is the number of events predicted in $B_{l4}$ decay for $B^\pm$
meson, $N$ is the number of $B$-mesons produced,
and $Br$ is the branching fraction of the relevant $B$ decay mode. 
For a realistic detection efficiency
$\e$, we have to rescale the number of events by this parameter,
$N_-+N_+\to\e (N_-+N_+)$. Taking $N_{SD}=1$, we obtain 
the number $N_B$ of the B mesons needed to observe CP violation at $1$-$\sigma$ level:
\be
N_B=\frac{1}{Br\cdot A^2}.
\ee

Next, we consider the so-called optimal observable.
An appropriate real weight function $w(s_{_M},q^2;\theta^*,\theta_l,\phi^*)$
is usually employed to separate the CP-odd ${\cal D}$ contribution and to enhance
its analysis power for the CP-odd parameter through 
the CP-odd quantity:
\begin{eqnarray}
\langle w{\cal D}\rangle\equiv\int\left[w{\cal D}\right] d\Phi_4,
\end{eqnarray}
and the analysis power is determined by the parameter,
\begin{eqnarray}
\varepsilon
   =\frac{\langle w{\cal D}\rangle}{\sqrt{\langle{\cal S}\rangle
          \langle w^2{\cal S}\rangle}}\;.
\label{Significance}
\end{eqnarray}
For the analysis power $\varepsilon$, the number $N_B$ of the $B$-mesons 
needed to observe CP violation at 1-$\sigma$ level is
\begin{eqnarray}
N_B=\frac{1}{Br\cdot\varepsilon^2}\;.
\label{eq:number}
\end{eqnarray}
Certainly, it is desirable to find the optimal weight function
with the largest analysis power. It is known  \cite{Optimal} that 
when the CP-odd contribution to the total rate is relatively small, 
the optimal weight function  is approximately given as
\begin{eqnarray}
w_{\rm opt}(s_{_M},s_L;\theta,\theta_l,\phi)=
 \frac{{\cal D}}{{\cal S}}~~~\Rightarrow~~~
 \varepsilon_{\rm opt}=\sqrt{\frac{\langle\frac{{\cal D}^2}{{\cal S}}\rangle}
 {\langle{\cal S}\rangle}}.
\end{eqnarray}
We adopt this optimal weight function in the following numerical analyses.

\section{Numerical Results and Conclusions}

\noindent
Now we show our numerical results. 
We use the so-called ISGW predictions \cite{ISGW}
for all the form factors in $B\to M_i$ transition amplitudes of Eq.~(\ref{BMamp}).
One can find in  Ref.~\cite{ISGW} the detailed description of the general 
formalism and relevant form factors for $B\to Xe\bar{\nu}_e$ 
after neglecting lepton masses.
In Appendix A, we give explicit expressions of form factors needed for 
semileptonic decays with non-zero lepton masses.

\begin{table}
{Table~2}. {The CP-violating rate asymmetry A and the optimal asymmetry
$\varepsilon_{\rm opt}$, determined within the SM,  
and the number of charged B meson pairs, 
$N_B$, needed for detection at $1\sigma$ level,
at reference value $\Im(\xi)=12.5$.}\par
\begin{tabular}{c|cc|cc|cc}
\hline
\multicolumn{7}{c}{$B\to \pi^+\pi^-l\bar{\nu}_l$}\\
\hline
Mode
&\multicolumn{2}{c|}{$l=e$}&\multicolumn{2}{c|}{$l=\mu$}&\multicolumn{2}{c}{$l=\tau$}\\
\hline\hline 
Asym. & Size(\%) & $N_B$ & Size(\%) & $N_B$ & Size(\%) & $N_B$\\
\hline
A & $0.94\times 10^{-6}$ & $1.37\times 10^{18}$ 
& $1.71\times 10^{-6}$ & $4.16\times 10^{17}$ & $1.14\times 10^{-6}$ & $1.46\times 10^{18}$\\
$\varepsilon_{\rm opt}$ & $1.45\times 10^{-2}$ & $5.75\times 10^{9}$ 
& $1.44\times 10^{-2}$ & $5.79\times 10^{9}$ & $1.11\times 10^{-2}$ & $1.56\times 10^{10}$\\
\hline
\end{tabular}
\end{table}

We first consider the case within the SM.
Total branching ratio of the $B^-\to (\sum_i M_i \to \pi^+\pi^-) e\bar{\nu}_e,~ 
M_i=\rho, f_{0,2}, D$ is about $0.8\%$.
It depends on the chosen value for $|V_{ub}|$, and
here we adopt the result by CLEO \cite{CLEO}
\be
|V_{ub}|=3.3\pm 0.4 \pm 0.7\times 10^{-3}.
\ee
In Table 2, we show the results of $B\to \pi\pi l\nu$ decays 
for the two CP-violating asymmetries;
the rate asymmetry $A$ and the optimal asymmetry $\varepsilon_{\rm opt}$.
We estimated the number of $B$-meson pairs, $N_B$, needed for detection at 
$1\sigma$ level for maximally-allowed values of CP-odd parameters $\Im(\xi)$ in
Eq.~(\ref{xi}). We use the current experimental bound \cite{pdg}
\be
\left| \frac{V_{ub}}{V_{cb}}\right|=0.08\pm 0.02,
\ee
which means
\be
|\xi|=12.5\pm 3.13.
\ee
The results in Table 2 are for the maximal case with $\Im(\xi)=|\xi|=12.5$.
Due to the large cancelations in the simple rate asymmetry \cite{cancel}
when we integrated over the phase space,
the optimal observable gives much better result. 
For example, using the optimal observable, we need  $\sim 10^9$  
$B$-meson pairs to detect the maximal CP-odd effect in electron mode.
CP violation effects in $B\to \pi^+\pi^-l\bar{\nu}_l$ decays within the SM 
are not likely to be detected, with ${\cal O}(10^8)$ $B$-meson pairs to be produced
at the asymmetric $B$ factories.  
One may rely on hadronic $B$-factories of BTeV and LHC-B.

\begin{table}
{Table~3}. {The CP-violating rate asymmetry A and the optimal asymmetry
$\varepsilon_{\rm opt}$, determined in the extended models,  
and the number of charged B meson pairs, 
$N_B$, needed for detection at $1\sigma$ level,
at reference values 
(a) $\Im(\zeta_{_{MHD}})=2.06$, $\Im(\zeta^{^I}_{_{SLQ}})=2.76$ and 
$\Im(\zeta^{^{II}}_{_{SLQ}})=18.4$ for the $B_{\tau 4}$ decays
and (b) $\Im(\zeta_{_{MHD}})=0.12$, $\Im(\zeta^{^I}_{_{SLQ}})=0.37$ and 
$\Im(\zeta^{^{II}}_{_{SLQ}})=1.84$ for the $B_{\mu 4}$ decays.}\par
\begin{tabular}{c|cc|cc|cc}
\hline
\multicolumn{7}{c}{(a) $B^-\to \pi^+\pi^-\tau\bar{\nu}_\tau$ mode}\\
\hline
Model
&\multicolumn{2}{c|}{MHD}&\multicolumn{2}{c|}{SLQ I}&\multicolumn{2}{c}{SLQ II}\\
\hline\hline 
Asym. & Size(\%) & $N_B$ & Size(\%) & $N_B$ & Size(\%) & $N_B$\\
\hline
A & $1.47\times 10^{-3}$ & $7.63\times 10^{11}$ & $2.67\times 10^{-3}$ & $1.99\times 10^{11}$ 
  & $3.62\times 10^{-3}$ & $7.39\times 10^{9}$\\
$\varepsilon_{\rm opt}$ & $16.2$ & $6.23\times 10^3$ & $18.2$ & 
    $4.27\times 10^3$ & $9.67$ & $1.04\times 10^{3}$\\
\hline
\multicolumn{7}{c}{ }\\
\hline
\multicolumn{7}{c}{(b) $B^-\to \pi^+\pi^-\mu\bar{\nu}_\mu$ mode}\\
\hline
Model
&\multicolumn{2}{c|}{MHD}&\multicolumn{2}{c|}{SLQ I}&\multicolumn{2}{c}{SLQ II}\\
\hline\hline 
Asym. & Size(\%) & $N_B$ & Size(\%) & $N_B$ & Size(\%) & $N_B$\\
\hline
A & $2.61\times 10^{-5}$ & $1.93\times 10^{15}$ & $0.90\times 10^{-4}$ & $1.59\times 10^{14}$ 
  & $3.43\times 10^{-4}$ & $8.89\times 10^{12}$\\
$\varepsilon_{\rm opt}$ & $0.18$ & $3.89\times 10^7$ & $0.50$ & 
    $5.13\times 10^6$ & $1.48$ & $4.76\times 10^{5}$\\
\hline
\end{tabular}
\end{table}

Next we consider the extended model case.
In this case, CP violation effects are proportional to the lepton mass,
and we consider only massive lepton ($\mu$ or $\tau$) cases.
In Table 3, we show the results of $B_{\tau 4}$ and $B_{\mu 4}$ decays.
Here in order to distinguish  new physics effect from the SM one,
we use a cutoff for the invariant mass of the final state $\pi^+\pi^-$  as
\be
\sqrt{s_{_M}}\le 1.4\;{\rm GeV}.
\ee
We consider only the lowest three
$u\bar{u}$ states, $\rho(770)$, $f_0(980)$ and $f_2(1275)$ 
as intermediate resonances in Table 1
so that the effects of $D$ meson can not enter, and  we can thus ensure
that the result is solely from new physics.
Similarly as in the SM case,
we estimate the number of $B$-meson pairs, $N_B$, needed for detection at 
$1\sigma$ level for {\it maximally-allowed} values of CP-odd parameters $\Im(\zeta)$ of
Eq.~(\ref{Hmodel}) and (\ref{Models}).
We again find the optimal observable gives much better results than 
the simple rate asymmetry. 

The results in Table 3 show that CP violation effects from new physics 
are readily observed in the forthcoming asymmetric $B$-factories, 
by using optimal observables.
As expected,
$B_{\tau 4}$ decay modes give better results than $B_{\mu 4}$ cases
for the MHD model, where the CP-odd parameter itself is proportional to 
the lepton mass.
For example, the current bounds in the MHD model
$$\Im(\zeta_{_{MHD}})=2.06\;(0.12)~~~{\rm for}~~~ \tau\;(\mu)$$ 
directly result from the lepton mass dependence.
But there is no such dependence in the SLQ models.
The current numerical values of CP-odd parameters in the SLQ models,
\be
\Im(\zeta^{^I}_{_{SLQ}})&=&2.76\;(0.37) \nonumber\\ 
\Im(\zeta^{^{II}}_{_{SLQ}})&=&18.4\;(1.84)~~~{\rm for}~~~ \tau\;(\mu), \nonumber
\ee
are just from different experimental bounds. 
Therefore, the smaller CP-odd value for $\mu$ family is a consequence of the fact 
that the current experimental constraints on the muon mode are more stringent.
And $B_{\tau 4}$ decay modes would provide more stringent constraints 
to all the extended models that we have considered. 

In conclusion, we have investigated
direct CP violations from physics beyond the SM as well as within the SM
through semileptonic $B_{l4}$
decays: $B^\pm\to \pi^+\pi^- l^\pm\nu_l$. 
Within the SM, CP violation could be generated through interference between
resonances with different quark flavors, that is, with different CKM matrix
elements.
We included $u\bar{u}$ state mesons $(\rho, f_0\;{\rm and}\;f_2)$ and
$D$ meson as intermediate resonances which decay to $\pi^+\pi^-$.
Using optimal observables, we found ${\cal O}(10^9)$  $B$-meson pairs are 
needed to probe CP violation effects at $1\sigma$ level
for the current maximal value of $\Im(\xi)=|V_{cb}/V_{ub}|=12.5$.
We have also investigated CP violation effects in extensions of the SM.
We considered multi-Higgs doublet model and scalar-leptoquark models.
Here CP violation is implemented 
through interference between $W$-exchange diagrams and 
scalar-exchange diagrams with complex couplings in the extended models.
We calculated the CP-odd rate asymmetry and the optimal asymmetry 
for $B_{\tau 4}$ and $B_{\mu 4}$ decay modes. 
We found that the optimal asymmetries for both modes are sizable and
can be detected at $1\sigma$ level with about $10^3$-$10^7$  
$B$-meson pairs, 
for maximally-allowed values of CP-odd parameters.
Since $\sim$$10^8$ $B$-meson pairs are expected to be produced yearly 
at the asymmetric $B$ factories,
one could easily investigate CP-violation effects in these decay modes $B_{l4}$
to extract much more stringent constraints on CP-odd parameters, 
$\Im(\zeta_{_{MHD}})$ and $\Im(\zeta^{^{I,II}}_{_{SLQ}})$.


\section*{Acknowledgments}

\noindent
We thank G. Cvetic for careful reading of the manuscript and 
his valuable comments.
The work of C.S.K. was supported 
in part by KRF Non-Directed-Research-Fund, Project No. 1997-001-D00111,
in part by the BSRI Program, Ministry of Education, Project No. 98-015-D00061,
in part by the KOSEF-DFG large collaboration project, 
Project No. 96-0702-01-01-2.
J.L. and W.N. wish to acknowledge the financial support of 
1997-sughak program of Korean Research Foundation.


\newpage
\noindent
{\Large\bf Appendix}


\begin{appendix}

\section{Form factors}

Form factors in Eq.~(\ref{BMamp}) within ISGW model \cite{ISGW} are
\be
u_+(q^2)=-F_5(q^2;f_0)\frac{m_u m_b m_q}{\sqrt{6}\beta_B \tilde{m}_{f_0}\mu_-},\;\;
u_+(q^2)=F_5(q^2;f_0)\frac{m_u(\tilde{m}_B+\tilde{m}_{f_0})}{\sqrt{6}\beta_B\tilde{m}_{f_0}}
\ee
\be
&&g(q^2)=\frac{F_3(q^2;\rho)}{2}\left[\frac{1}{m_q}-\frac{m_u \beta^2_B}
                      {2\mu_-\tilde{m}_\rho \beta^2_{B\rho}}\right],\;\;
  f(q^2)=2\tilde{m}_B F_3(q^2;\rho),\nonumber\\
&&a_+(q^2)=-\frac{F_3(q^2;\rho)}{2\tilde{m}_\rho}\left[1+\frac{m_u}{m_b}
            \left(\frac{\beta_B^2-\beta_\rho^2}{\beta_B^2+\beta_\rho^2}\right)
             -\frac{m_u^2\beta_\rho^2}{4\mu_-\tilde{m}_B\beta_{B\rho}^4}\right],\nonumber\\
&&a_-(q^2)=\frac{F_3(q^2;\rho)}{2\tilde{m}_\rho}\left[1+\frac{m_u}{m_b}
            \left(1+\frac{m_u\beta_\rho^2}{m_q\beta_{B\rho^2}}\right)
                 -\frac{m_u^2\beta_\rho^2}{4\mu_+\tilde{m}_B\beta_{B\rho}^4}\right]
\ee
\be
&&h(q^2)=F_5(q^2;f_2)\frac{m_u}{2\sqrt{2}\tilde{m}_B\beta_B}
            \left[\frac{1}{m_q}-\frac{m_u \beta^2_B}
                      {2\mu_-\tilde{m}_{f_2} \beta^2_{Bf_2}}\right],\;\;
  k(q^2)=\sqrt{2}\frac{m_u}{\beta_B} F_5(q^2;f_2),\nonumber\\
&&b_+(q^2)=-\frac{F_5(q^2;f_2)m_u}{2\sqrt{2}\beta_B\tilde{m}_{f_2}m_b}\left[1
               -\frac{m_u\beta_{f_2}^2}{\tilde{m}_B\beta_{Bf_2}^2}
        -\frac{m_u^2m_b\beta_{f_2}^4}{4\mu_-\tilde{m}_B^2\beta_{Bf_2}^4}\right],\nonumber\\
&&b_-(q^2)=\frac{F_5(q^2;f_2)m_u}{2\sqrt{2}\beta_B\tilde{m}_{f_2}m_b}\left[1
               +\frac{m_u^2\beta_{f_2}^2}{m_q\tilde{m}_B\beta_{Bf_2}^2}
               -\frac{m_u^2m_b\beta_{f_2}^4}{4\mu_+\tilde{m}_B^2\beta_{Bf_2}^4}\right]
\ee
\be
&&f_+(q^2)=F_3(q^2;D)\left[1+\frac{m_b}{2\mu_-}-\frac{m_bm_qm_u\beta_B^2}
               {4\mu_+\mu_-\tilde{m}_D\beta_{BD}^2}\right],\nonumber\\
&&f_-(q^2)=F_3(q^2;D)\left[1-(\tilde{m}_B+\tilde{m}_D)
            \left(\frac{1}{2m_q}-\frac{m_u\beta_B^2}{4\mu_+\tilde{m}_D\beta_{BD}^2}\right)
            \right]
\ee
where
\be
\beta_{BX}^2=\frac{1}{2}(\beta_B^2+\beta_X^2),\;\;
\mu_\pm=\left(\frac{1}{m_q}\pm\frac{1}{m_b}\right)^{-1}.
\ee
And
\be
F_n(q^2;X)=\left(\frac{\tilde{m}_X}{\tilde{m}_B}\right)^{1/2}
           \left(\frac{\beta_B\beta_X}{\beta_{BX}^2}\right)^{n/2}
           \exp\left[-\left(\frac{m_u^2}{4\tilde{m}_B\tilde{m}_X}\right)
                     \frac{q_m-q^2}{\kappa^2\beta_{BX}^2}\right],
\ee
where relativistic compensation factor $\kappa=0.7$, and $q_m$ is the maximum value
of $q^2$:
\be
q_m=(m_B-\sqrt{s_{_M}})^2,
\ee
and $m_q$ is $m_u$ for $u\bar{u}$ state mesons and $m_c$ for $D$-mesons.
The numerical values of $\beta_X$ in GeV unit are
\be
\beta_B=0.41,\;\beta_D=0.39,\;\beta_{f_0}=0.27,\;\beta_\rho=0.31,\;\beta_{f_2}=0.27,
\ee
and quark masses in GeV unit are
\be
m_u=0.33,\; m_c=1.82,\;m_b=5.12.
\ee
The so-called mock meson masses $\tilde{m}_X$ are defined as
\be
\tilde{m}_B=m_b+m_u,\; \tilde{m}_D=m_c+m_u,\; \tilde{m}_{\rho,f_0,f_2}=2m_u.
\ee
\section{Kinematics}

\begin{center}
{\bf Spherical harmonics}
\end{center}

\be
&&Y^0_0=\tilde{Y}^0_0=\frac{1}{\sqrt{4\pi}},\nonumber\\
&&Y_1^0=\sqrt{\frac{3}{4\pi}}\cos\theta,\; 
       Y_1^{\pm 1}=\mp\sqrt{\frac{3}{8\pi}}\sin\theta e^{\pm i\phi},\nonumber\\
&&Y_2^0=\sqrt{\frac{5}{4\pi}}(\frac{3}{2}\cos^2\theta-\frac{1}{2}),\; 
       Y_2^{\pm 1}=\mp\sqrt{\frac{15}{8\pi}}\sin\theta\cos\theta e^{\pm i\phi},
\ee

\begin{center}
{\bf Polarization vectors}
\end{center}

\noindent
In the $B$ rest frame, where the coordinates are chosen such that the $z$-axis is along the 
$M_i$ momentum and the charged lepton momentum is in the $x$--$z$ plane with positive
$x$-component (cf. Fig.~2a),
the polarization vectors for the virtual $W$ are
\be
\e(q,\pm)^\mu&=&\mp\frac{1}{\sqrt{2}}(0,1,\mp i,0),\nonumber\\
\e(q,0)^\mu&=&\frac{1}{\sqrt{q^2}}(p_M,0,0,-q^0),\nonumber\\
\e(q,s)^\mu&=&\frac{1}{\sqrt{q^2}}q^\mu,
\ee
and the polarization states of the spin 1 mesons are
\be
\e(\pm 1)^\mu=\mp\frac{1}{\sqrt{2}}(0,1,\pm i,0),\;
\e(0)^\mu=\frac{1}{\sqrt{s_{_M}}}(p_M,0,0,E_M),
\ee
where $p_M=\sqrt{Q_+Q_-}/2m_B$ with $Q_\pm$ defined in Eq.~(\ref{Qpm}),
and $E_M=(m_B^2+s_{_M}-q^2)/2m_B$. 
And for the spin 2 meson we get
\be
\e(\pm 2)^{\mu\nu}&=&\e^\mu(\pm 1)\e^\nu(\pm 1),\nonumber\\
\e(\pm 1)^{\mu\nu}&=&\frac{1}{\sqrt{2}}\left[\e^\mu(\pm 1)\e^\nu(0)
                     +\e^\mu(0)\e^\nu(\pm 1)\right],\nonumber\\
\e(0)^{\mu\nu}&=&\frac{1}{\sqrt{6}}\left[\e^\mu(+1)\e^\nu(-1)
                     +\e^\mu(-1)\e^\nu(+1)\right]+\sqrt{\frac{2}{3}}\e^\mu(0)\e^\nu(0).
\ee

In the $W$ rest frame the polarization states of the virtual $W$ are
\be
\e(q,\pm)^\mu&=&\mp\frac{1}{\sqrt{2}}(0,1,\mp i,0),\nonumber\\
\e(q,0)^\mu&=&(0,0,0,-1),\nonumber\\
\e(q,s)^\mu&=&\frac{1}{\sqrt{q^2}}q^\mu=(1,0,0,0).
\ee

\section{CP-even and CP-odd quantities}

\begin{center}
{\bf Within the SM}
\end{center}

\noindent
The CP-even quantity ${\cal S}$ is
\be
{\cal S}=2C(q^2,s_{_M})\Sigma,
\ee
with
\be
\Sigma&=&(L^-_0S^0_0Y^0_0)^2|\Pi_{f_0}|^2+(L^-_0P^0_0Y^0_0)^2|\xi|^2|\Pi_D|^2+|\langle V^-\rangle \Pi_\rho|^2+|\langle T^-\rangle \Pi_{f_2}|^2\nonumber\\
  &&+2(L^-_0S^0_0Y^0_0)\Re(\Pi_{f_0}\Pi_\rho^*\langle V^-\rangle^*+\Pi_{f_0}\Pi_{f_2}\langle T^-\rangle^*)+2\Re(\Pi_\rho\Pi_{f_2}^*\langle V^-\rangle \langle T^-\rangle^*)\nonumber\\
  &&+2(L^-_0P^0_0Y^0_0)\Re(\xi)[(L^-_0S^0_0Y^0_0)\Re(\Pi_D\Pi_{f_0}^*)+\Re(\Pi_D\Pi_\rho^*\langle V^-\rangle^*+\Pi_D\Pi_{f_2}^*\langle T^-\rangle^*)]\nonumber\\
  &&+(L^+_0S^0_0Y^0_0-L^+_sS^0_sY^0_0)^2|\Pi_{f_0}|^2+(L^+_0P^0_0Y^0_0-L^+_sP^0_sY^0_0)^2|\xi|^2|\Pi_D|^2\nonumber\\
  &&+|\Pi_\rho|^2|\langle V^+\rangle -L^+_sV^0_sY^0_1|^2+|\Pi_{f_2}|^2|\langle T^+\rangle -L^+_sT^0_sY^0_2|^2\nonumber\\
  &&+2(L^+_0S^0_0-L^+_sS^0_s)Y^0_0[-(L^+_sV^0_sY^0_1)\Re(\Pi_{f_0}\Pi_\rho^*)+\Re(\Pi_{f_0}\Pi_\rho^*\langle V^+\rangle^*)\nonumber\\
  &&               -(L^+_sT^0_sY^0_2)\Re(\Pi_{f_0}\Pi_{f_2}^*)+\Re(\Pi_{f_0}\Pi_{f_2}^*\langle T^+\rangle^*)]\nonumber\\
  &&+2\Re(\Pi_\rho\Pi_{f_2}^*\langle V^+\rangle \langle T^+\rangle^*)+2(L^+_sV^0_sY^0_1)(L^+_sT^0_sY^0_2)\Re(\Pi_\rho\Pi_{f_2}^*)\nonumber\\
  &&-2(L^+_sT^0_sY^0_2)\Re(\Pi_\rho\Pi_{f_2}^*\langle V^+\rangle )-2(L^+_sV^0_sY^0_1)\Re(\Pi_\rho\Pi_{f_2}^*\langle T^+\rangle^*)\nonumber\\
  &&+2(L^+_0P^0_0-L^+_sP^0_s)Y^0_0\Re(\xi)[(L^+_0S^0_0-L^+_sS^0_s)Y^0_0\Re(\Pi_D\Pi_{f_0}^*)
                    -(L^+_sV^0_sY^0_1)\Re(\Pi_D\Pi_\rho^*)\nonumber\\
  &&       +\Re(\Pi_D\Pi_\rho^*\langle V^+\rangle^*)-(L^+_sT^0_sY^0_2)\Re(\Pi_D\Pi_{f_2}^*)+\Re(\Pi_D\Pi_{f_2}^*\langle T^+\rangle^*)],
\ee
and the CP-odd quantity ${\cal D}$ is
\be
{\cal D}=2\Im(\xi)C(q^2,s_{_M})\Delta,
\ee
with
\be
\Delta&=&-2(L^-_0P^0_0Y^0_0)[(L^-_0S^0_0Y^0_0)\Im(\Pi_D\Pi_{f_0}^*)+\Im(\Pi_D\Pi_\rho^*\langle V^-\rangle^*)+\Im(\Pi_D\Pi_{f_2}^*\langle T^-\rangle^*)]\nonumber\\
    &&-2(L^+_0P^0_0-L^+_sP^0_s)Y^0_0[(L^+_0S^0_0-L^+_sS^0_s)Y^0_0\Im(\Pi_D\Pi_{f_0}^*)-(L^+_sV^0_sY^0_1)\Im(\Pi_D\Pi_\rho^*)\nonumber\\
    &&+\Im(\Pi_D\Pi_\rho^*\langle V^+\rangle^*)-(L^+_sT^0_sY^0_2)\Im(\Pi_D\Pi_{f_2}^*)+\Im(\Pi_D\Pi_{f_2}^*\langle T^+\rangle^*)], 
\ee
where
\be
\langle V^\pm\rangle \equiv\sum_{i=0,\pm 1}L^\pm_\l V^\l_\l Y^\l_1,\;\;
\langle T^\pm\rangle \equiv\sum_{i=0,\pm 1}L^\pm_\l T^\l_\l Y^\l_2,
\ee
and the overall function $C(q^2,s_{_M})$ is given by
\be
C(q^2,s_{_M})=|V_{ub}|^2\frac{G_F^2}{2}\frac{1}{2m_B}
       \frac{(q^2-m_l^2)\sqrt{Q_+Q_-}}{256\pi^3m_B^2q^2}.
\ee

\begin{center}
{\bf With a complex scalar coupling}
\end{center}

\noindent
The CP-even quantity ${\cal S}$ is
\be
{\cal S}=2C(q^2,s_{_M})\Sigma,
\ee
with
\be
\Sigma&=&(L^-_0S^0_0Y^0_0)^2|\Pi_{f_0}|^2+|\langle V^-\rangle \Pi_\rho|^2+|\langle T^-\rangle \Pi_{f_2}|^2\nonumber\\
   &&+2(L^-_0S^0_0Y^0_0)\Re(\Pi_{f_0}\Pi_\rho^*\langle V^-\rangle^*+\Pi_{f_0}\Pi_{f_2}^*\langle T^-\rangle^*)+2\Re(\Pi_\rho\Pi_{f_2}^*\langle V^-\rangle \langle T^-\rangle^*)\nonumber\\
   &&+|\Pi_{f_0}|^2|L^+_0S^0_0Y^0_0-(1-\zeta^\prime)L^+_sS^0_sY^0_0|^2\nonumber\\
   &&+|\Pi_\rho|^2[|\langle V^+\rangle|^2 +(L^+_sV^0_sY^0_1)^2|1-\zeta^\prime|^2
      -2(L^+_sV^0_sY^0_1)\Re(\langle V^+\rangle)\Re(1-\zeta^\prime)]\nonumber\\
   && +|\Pi_{f_2}|^2[|\langle T^+\rangle|^2 +(L^+_sT^0_sY^0_2)^2|1-\zeta^\prime|^2
      -2(L^+_sT^0_sY^0_2)\Re(\langle T^+\rangle)\Re(1-\zeta^\prime)]\nonumber\\
   &&+2\Re(\Pi_{f_0}\Pi_\rho^*)[(L^+_0S^0_0-L^+_sS^0_s)Y^0_0\Re(\langle V^+\rangle )-(L^+_0S^0_0Y^0_0)(L^+_sV^0_sY^0_1)\Re(1-\zeta^\prime)\nonumber\\
   &&       +(L^+_sS^0_sY^0_0)\Re(\langle V^+\rangle )\Re(\zeta^\prime)+(L^+_sS^0_sY^0_0)(L^+_sV^0_sY^0_1)|1-\zeta^\prime|^2]\nonumber\\
   &&+2\Im(\Pi_{f_0}\Pi_\rho^*)\Im(\langle V^+\rangle )[(L^+_0S^0_0-L^+_sS^0_s)Y^0_0+(L^+_sS^0_sY^0_0)\Re(\zeta^\prime)]\nonumber\\
   &&+2\Re(\Pi_{f_0}\Pi_{f_2}^*)[(L^+_0S^0_0-L^+_sS^0_s)Y^0_0\Re(\langle T^+\rangle )-(L^+_0S^0_0Y^0_0)(L^+_sT^0_sY^0_2)\Re(1-\zeta^\prime)\nonumber\\
   &&       +(L^+_sS^0_sY^0_0)\Re(\langle T^+\rangle )\Re(\zeta^\prime)+(L^+_sS^0_sY^0_0)(L^+_sT^0_sY^0_2)|1-\zeta^\prime|^2]\nonumber\\
   &&+2\Im(\Pi_{f_0}\Pi_{f_2}^*)\Im(\langle T^+\rangle )[(L^+_0S^0_0-L^+_sS^0_s)Y^0_0+(L^+_sS^0_sY^0_0)\Re(\zeta^\prime)]\nonumber\\
   &&+2\Re(\Pi_\rho\Pi_{f_2}^*)[\Re(\langle V^+\rangle \langle T^+\rangle^*)-(L^+_sT^0_sY^0_2)\Re(\langle V^+\rangle )+(L^+_sT^0_sY^0_2)\Re(\langle V^+\rangle )\Re(\zeta^\prime)\nonumber\\
   &&             -(L^+_sV^0_sY^0_1)\Re(\langle T^+\rangle )\Re(1-\zeta^\prime)+(L^+_sV^0_sY^0_1)(L^+_sT^0_sY^0_2)|1-\zeta^\prime|^2]\nonumber\\
   &&-2\Im(\Pi_\rho\Pi_{f_2}^*)[\Im(\langle V^+\rangle \langle T^+\rangle^*)-(L^+_sT^0_sY^0_2)\Im(\langle V^+\rangle )+(L^+_sT^0_sY^0_2)\Im(\langle V^+\rangle )\Re(\zeta^\prime)\nonumber\\
   &&             +(L^+_sV^0_sY^0_1)\Im(\langle T^+\rangle )\Re(1-\zeta^\prime)],\nonumber\\
\ee
and the CP-odd quantity ${\cal D}$ is
\be
{\cal D}=2\Im(\zeta^\prime)C(q^2,s_{_M})\Delta,
\ee
with
\be
\Delta&=&2\bigg[
   \Im(\langle V^+\rangle )\{(L^+_sV^0_sY^0_1)|\Pi_\rho|^2+(L^+_sS^0_sY^0_0)\Re(\Pi_{f_0}\Pi_\rho^*)+(L^+_sT^0_sY^0_2)\Re(\Pi_\rho\Pi_{f_2}^*)\}\nonumber\\
&&+\Im(\langle T^+\rangle )\{(L^+_sT^0_sY^0_2)|\Pi_{f_2}|^2+(L^+_sS^0_sY^0_0)\Re(\Pi_{f_0}\Pi_{f_2}^*)+(L^+_sV^0_sY^0_1)\Re(\Pi_\rho\Pi_{f_2}^*)\}\nonumber\\
&&+\Re(\langle V^+\rangle )\{(L^+_sT^0_sY^0_2)\Im(\Pi_\rho\Pi_{f_2}^*)-(L^+_sS^0_sY^0_0)\Im(\Pi_{f_0}\Pi_\rho^*)\}\nonumber\\
&&-\Re(\langle T^+\rangle )\{(L^+_sV^0_sY^0_1)\Im(\Pi_\rho\Pi_{f_2}^*)+(L^+_sS^0_sY^0_0)\Im(\Pi_{f_0}\Pi_{f_2}^*)\}\nonumber\\
&&+(L^+_0S^0_0Y^0_0)(L^+_sV^0_sY^0_1)\Im(\Pi_{f_0}\Pi_\rho^*)+(L^+_0S^0_0Y^0_0)(L^+_sT^0_sY^0_2)\Im(\Pi_{f_0}\Pi_{f_2}^*)\bigg].
\ee
Note that since every term in $\Delta$ of Eq. (90) contains square terms of
$L^+_i$ which are proportional to $m_l$ (see Eq.~(\ref{Lpm})),
the CP-odd quantity ${\cal D}$ of Eq. (89)
is proportional to lepton mass
due to the definition of $\zeta^\prime$ (see Eq. (\ref{zetap})).

\end{appendix}
%

\end{document}